\documentclass[prd,aps,twocolumn,showpacs,floats,floatfix]{revtex4}
\usepackage{graphicx}
\usepackage{amsmath}
\usepackage{subfigure}
\usepackage{appendix}
\usepackage[normalem]{ulem}
\usepackage[utf8]{inputenc}
\usepackage{mathrsfs}
\usepackage{latexsym}
\usepackage{algorithm}
\usepackage{algpseudocode}
\usepackage{amsmath}

\usepackage{graphicx}
\usepackage{subfigure}
\usepackage{parskip}
\usepackage{dcolumn}
\usepackage{bm}
\usepackage{amssymb}
\usepackage{latexsym}
\usepackage{ulem}
\usepackage[colorlinks,linkcolor=magenta,anchorcolor=cyan,citecolor=blue]{hyperref}
\usepackage{color}
\usepackage{hyperref}
\usepackage[hyphenbreaks]{breakurl}
\usepackage{url}

\def\be{\begin{equation}}
\def\ee{\end{equation}}
\def\ba{\begin{eqnarray}}
\def\ea{\end{eqnarray}}

\usepackage{color}
\usepackage{acronym}
\usepackage{multirow}
\usepackage{tabularx}
\usepackage{hyperref}
\usepackage{microtype}
\usepackage{graphicx}
\usepackage{subfigure}
\usepackage{booktabs} 
\usepackage[sort&compress]{natbib}
\usepackage{amssymb}
\usepackage{mathtools}
\usepackage{bm}
\usepackage{cancel}
\usepackage{listings}
\begin{document}
\title{Self-supervised learning for gravitational wave signal identification}

\author{Hao-Yang Liu$^{3} $\footnote{\href{liuhaoyang19@mails.ucas.ac.cn}{liuhaoyang19@mails.ucas.ac.cn}}}
\author{Yu-Tong Wang$^{1,2} $\footnote{\href{wangyutong@ucas.ac.cn}{wangyutong@ucas.ac.cn}}}

\affiliation{$^1$ International Center for Theoretical Physics
        Asia-Pacific, Beijing/Hangzhou, China}

\affiliation{$^2$ School of Fundamental Physics and Mathematical
        Sciences, Hangzhou Institute for Advanced Study, UCAS, Hangzhou
        310024, China}

\affiliation{$^3$ School of Physics Sciences, University of
        Chinese Academy of Sciences, Beijing 100049, China}

\begin{abstract}

The computational cost of searching for gravitational wave (GW)
signals in low latency has always been a matter of concern. We
present a self-supervised learning model applicable to the GW
detection. Based on simulated massive black hole binary signals in
synthetic Gaussian noise representative of space-based GW
detectors Taiji and LISA sensitivity, and regarding their
corresponding datasets as a GW twins in the contrastive learning
method, we show that the self-supervised learning may be a highly
computationally efficient method for GW signal identification.

\end{abstract}

\maketitle
\setlength{\parindent}{1em}

\section{Introduction}

The field of gravitational wave (GW) detection has seen an
explosion of compact binary coalescence signals over the past
several years
\cite{LIGOScientific:2016aoc,LIGOScientific:2016sjg,LIGOScientific:2017bnn,LIGOScientific:2017vwq,LIGOScientific:2018mvr,LIGOScientific:2020ibl}.

Recently, 93 GW events have been reported
\cite{LIGOScientific:2021djp}.
It is expected that over the coming years, more GW events,
including binary black holes (BBH), binary neutron stars (BNS), as
well as other exotic sources will be observed more frequently. As
such, the need for more efficient search methods will be more
important as the detectors improve in sensitivity.

It is well-known that identifying signal is achieved, in part,
using a technique known as template-based matched filtering, which
uses a
bank~\cite{Brown:2012qf,DalCanton:2017ala,Harry:2009ea,DalCanton:2014hxh,Ajith:2012mn}
of template
waveforms~\cite{Sathyaprakash:1991mt,Taracchini:2013rva,Privitera:2013xza,Blanchet:2013haa}
spanning a large astrophysical parameter space.
The corresponding waveform models that cover the inspiral, merger,
and ringdown phases of a compact binary coalescence are based on
combining post-Newtonian
theory~\cite{Arun:2008kb,Buonanno:2009zt,Mishra:2016whh}, the
effective-one-body method~\cite{Buonanno:1998gg}, and numerical
relativity simulations~\cite{Pretorius:2005gq}. However, the
algorithms used by the search pipelines to make detections are
computationally expensive, since the large parameter space, as well as analysis of the high
frequency components of the waveform, inevitably result in large
computational cost.

Recently, the deep learning has been in popularity
\cite{Goodfellow:2014upx,Simonyan:2014cmh,Yuille:2014lcc,Fergus:2013mdz,Szegedy:2014nrf},
which is a subset of machine learning. The advantage of deep learning
is that the computationally intensive stage is pre-computed, so it
is able to performing analyses rapidly. Its successful
implementations include image
processing~\cite{Efros:2016aae,Karpathy:2014lff}, medical
diagnosis, in particular GW field, such as glitch
classification~\cite{Mukund:2016thr,Zevin:2016qwy,George:2017fbn}
and signal
identification~\cite{George:2016hay,Wei:2019zlc,Xia:2020vem,Lin:2020aps,Liao:2021vec,Ruan:2021fxq,Bacon:2022lsm,Zhao:2022qob,Schafer:2022dxv,Murali:2022sba,Nousi:2022dwh,Shen:2019vep,Ren:2022hny}.

However, the deep learning method currently used for searching for
the GW signals is often limited to building supervised learning
models, which are indeed trained on scarcely labelled or simulated
data and thus suffer from biases or small training sizes when
applied to new or full datasets. Thus it is significant to
consider the tailored deep learning models in a self-supervised,
semi-supervised or unsupervised
manner~\cite{Hinton:2002knh,Maaten:1912mlm}.

In this paper, we report the construction of a self-supervised
learning that can reproduce the searching for massive black hole
binary (MBHB) GW signals. A contrastive learning method, the GW
twins, which applies redundancy-reduction to self-supervised
learning~\cite{Deny:2103:mld}, is proposed. Based on simulated
MBHB signals in synthetic Gaussian noise representative of
space-based GW detectors Taiji \cite{Hu:2017mde} and LISA
\cite{LISA:2017pwj} sensitivity, we investigate whether the
self-supervised learning can efficiently identify a signal present
in the data, or the data contain only detector noise.

\section{Method}

\subsection{Description of GW twins }

\begin{figure*}
\includegraphics[width=18.7cm]{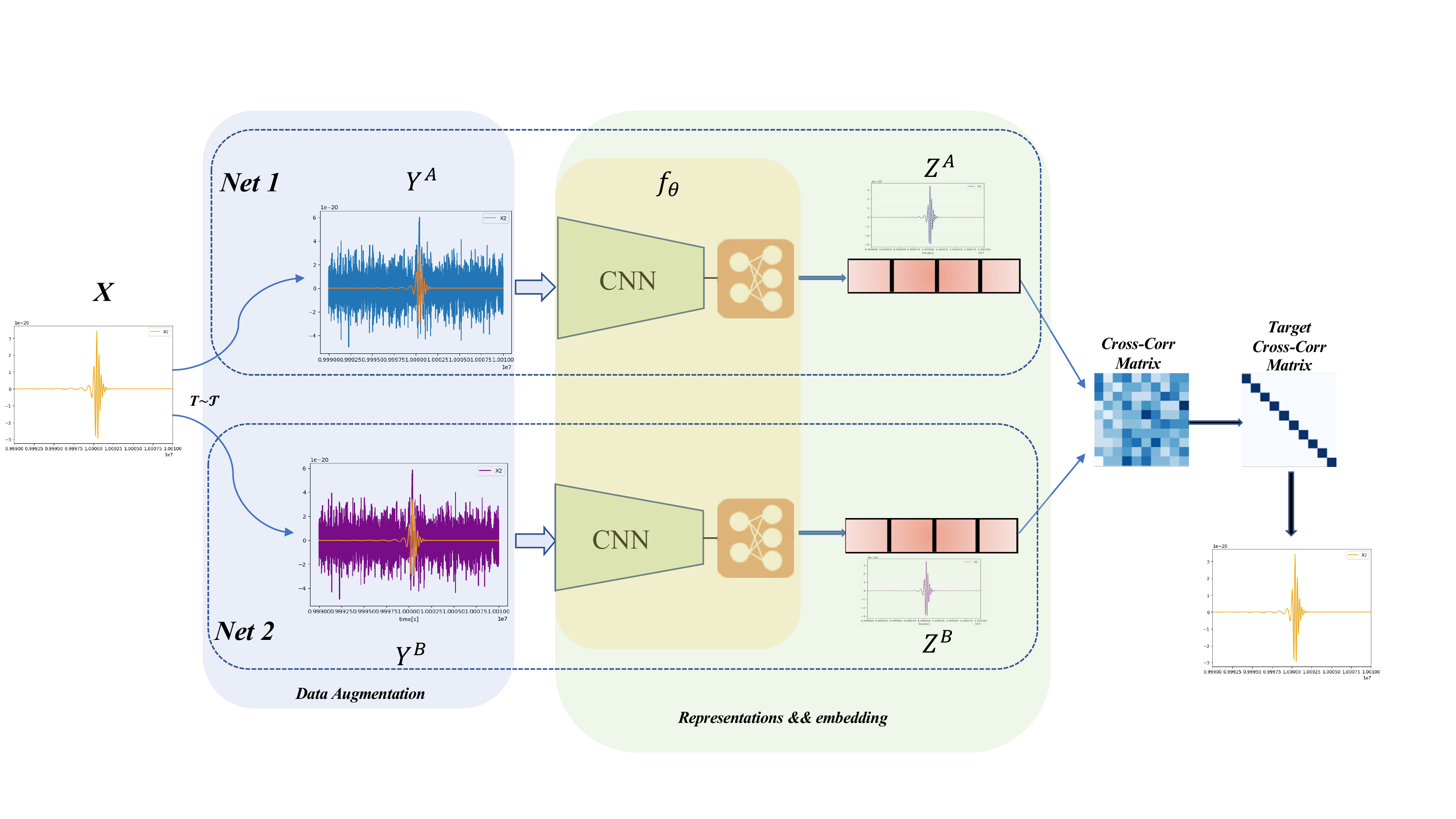}
\caption{The objective function of GW twins measures the
cross-correlation matrix between the embeddings of two identical
networks $h^A$ and $h^B$ (described as the extraction of GW signal
by Net A and Net B) fed with augmentation versions of a batch of
samples $d^A$ and $d^B$ (covered by the light blue area). Here,
$d^A=h + n_{L}$ and $d^B=h + n_{T}$, and the cross-correlation
matrix is made close to the identity. This causes the embedding
vectors of distorted versions of a sample (blue and purple signal
waveforms) to be similar, while minimizing the redundancy between
the components of the embedding vectors.\label{fig:fig_method}}
\end{figure*}

In self-supervised learning for GW detection, our GW twins
operates on a joint embedding of augmentation vectors, see
Fig.\ref{fig:fig_method}. In detail, it produces two augmentation
vectors for all data of a batch $h$ sampled from a dataset. In our
case, we consider the batch of $h$ as an injected GW waveform
signal. The augmentation vectors are obtained with a distribution
of data augmentations $\mathcal{T}$ (the light blue area in
Fig.\ref{fig:fig_method}). Two batches of augmentation vectors
$d^A$ and $d^B$ are fed to a deep network $f_{\theta}$ (covered by
the light yellow area in Fig.\ref{fig:fig_method}), producing
batches of embeddings $h^{A}$ and $h^{B}$, respectively.

In our GW twins design, the corresponding loss function
$\mathcal{L}_{GW}$ is
\begin{equation}
\mathcal{L}_{GW} =  \underbrace{\sum_i
\left(1-C^{GW}_{ii}\right)^2}_\text{invariant part}  + ~~\lambda
\underbrace{\sum_{i}\sum_{j \neq i}
\left({C^{GW}_{ij}}\right)^2,}_\text{redundancy reduction part}
\label{eq:lossBarlow}
\end{equation}
where
\begin{equation} C^{GW}_{ij} = \frac{ \sum_b h^A_{b,i}
h^B_{b,j}} {\sqrt{\sum_b {(h^A_{b,i})}^2} \sqrt{\sum_b
{(h^B_{b,j})}^2}} \label{eq:crosscorr}
\end{equation}
is the cross-correlation matrix computed between the outputs of
two identical networks, with the values comprised between -1 (i.e.
perfect anti-correlation) and 1 (i.e. perfect correlation),
$\lambda$ is a positive constant, $b$ is batch samples, and $i,j$
represent the vector dimension of the networks outputs

Here, the objective function (OF) is understood as an
instantiation of the information bottleneck (IB)
objective~\cite{Tishby:2000:tpb,Tishby:1503:tzy}.
It is easy to see that our OF has similarities with existing OFs
for self-supervised learning, for example, the redundancy
reduction part plays a role similar to the contrastive part in the
\textsc{infoNCE} objective~\cite{Vinyals:1807:olv}.

\subsection{Implementation Details}

It is often hard to find a large collection of correctly labelled
data. This scarcity of labelled real data exists for GW events as
well. One way to combat this shortage is to train on simulated
data and hope the trained model is useful for the real dataset as
well. The success of such an approach is dependent on the quality
of simulation. However, we will have rich raw data in GW
detection, specially for space-based detectors, such as Taiji and
LISA. Thus the application of self-supervised training can be
particularly relevant for GW signal detection.

The basic idea of our GW twins design for self-supervised training
is to have two dataset which see different versions of GW data and
use the OF of GW twins to learn embeddings. In details, we can
feed the data $d^A$ from LISA into Net A and the data $d^B$ from
Taiji into Net B. In Fig.\ref{fig:fig_method}, the blue strain and
purple strain in the data augmentation waveform plottings which
are covered by the light blue area correspond to the LISA noise
data strain $n_L$ and the Taiji noise data strain $n_T$, the
orange waveform is the GW signal $h$. Since the noise in Taiji is
different from LISA, we can imagine it as an augmented sample of
the same underlying signal.

It is well-known that most self-supervised works focused on
images, and so usually inputs are images and 2D CNN is used.
However, for faster training we will use a 1D CNN model inspired
by LISA and Taiji mock data challenge\cite{LISAMDC,TaijiMDC}. Moreover, we observed that adding a
GRU layer and reducing the original pooling size led to better
results. We also removed the final fully connected layers since we
require the models to produce embeddings. The bandpassed waveforms
are fed as input to the model and we get embeddings of size 2048
as output.

In the original paper for calculating $\mathcal{L}_{GW}$ in
(\ref{eq:lossBarlow}), they used LARS
optimizer~\cite{You:1708:ygg} but we found that AdamW with an
initial learning rate of $10^{-4}$ also works, though LARS
optimizer works better. Currently, we use LISA mock data
challenge and Taiji mock data challenge to generated noise as an
additional augmentation.

The details about the contrastive learning work flow are list as
follows:

\textit{Data augmentations} The input data is first converted to
produce the two augmentation vectors shown in
Fig.\ref{fig:fig_method}. The data augmentation pipeline includes:
random cropping, resizing, horizontal flipping, color jittering,
converting to grayscale, Gaussian blurring, and solarization. The
cropping and resizing are always applied, while the last five are
applied randomly, with some probability. This probability is
different for the two augmentation vectors in the last two
conversations (blurring and solarization). In this case we add the
different space-based detectors noise into the input data, this
data augmentation can be viewed as a color jittering. We use the
same augmentation parameters as \textsc{BYOL}.

\textit{Architecture} The encoder consists of a ResNet-50 network
(without the final classification layer, 2048 output units)
followed by a projector network. The projector network has three
linear layers, each with 8192 output units. The first two layers
of the projector are followed by a batch normalisation layer and
rectified linear units. We call the output of encoder the
``representations" and the output of projector the ``embeddings".
The representations are used for downstream tasks and the
embeddings are fed to $\mathcal{L}_{GW}$ of our GW twins.

\textit{Optimization} We follow the optimization protocol
described in \textsc{BYOL}. The LARS optimizer is used, and we
work for 1000 epochs with a batch size of 2048. We set the
learning rate 0.2 for the weights and 0.0048 for the biases and
batch normalization parameters. The learning rate warm-up period
of 10 epochs is adopted, after which we lowered the learning rate
by a factor of 1000 using a cosine decay schedule. The best
results of $\lambda$ in $\mathcal{L}_{GW}$ is $\lambda =
5\cdot10^{-3}$.

\section{Simulated space-based GW datasets}

\subsection{Time-delay interferometry and noise budgets}

Time-delay interferometry (TDI) will be employed for both LISA and
Taiji to suppress the laser frequency noise and achieve targeting
sensitivity. The principle of the TDI is to combine multiple
time-shifted interferometric links and obtain an equivalent equal
path for two interferometric laser beams. The GW response of TDI
is combined by the response of every single link. For a LISA-like
with six laser links, three optimal TDI channels (A, E, T) could
be constructed from three first-generation Michelson TDI
configuration (X, Y, Z), see
Refs.~\cite{Otto:2012dk,Otto:2015erp,Tinto:2018kij,Wang:2020pkk,Wang:2020fwa,Wang:2021mou,Prince:2002hp,Vallisneri:2007xa}
for details about the TDI configuration.

\begin{figure*}[tbp]
\includegraphics[width=0.5\textwidth]{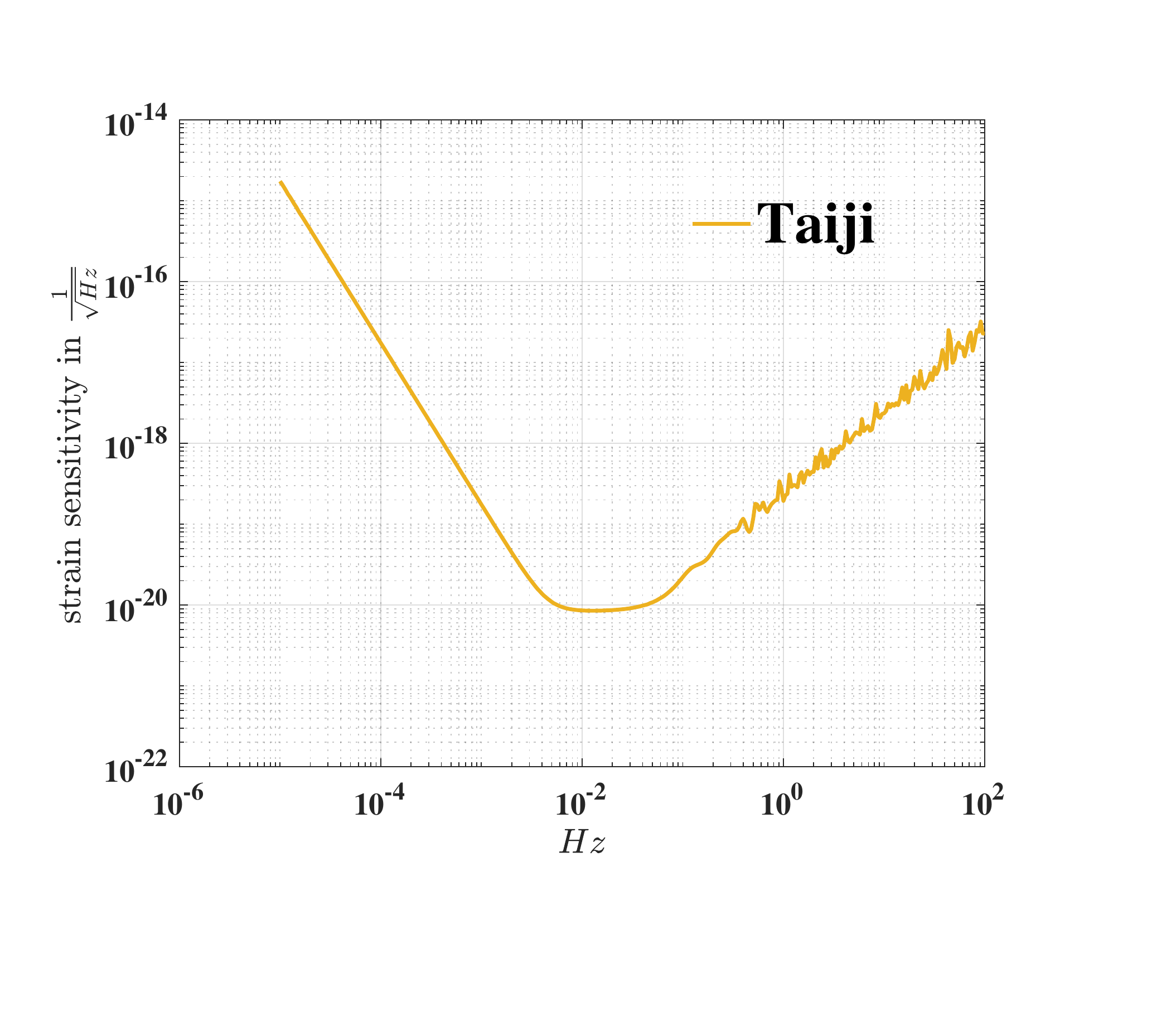}
\includegraphics[width=0.48\textwidth]{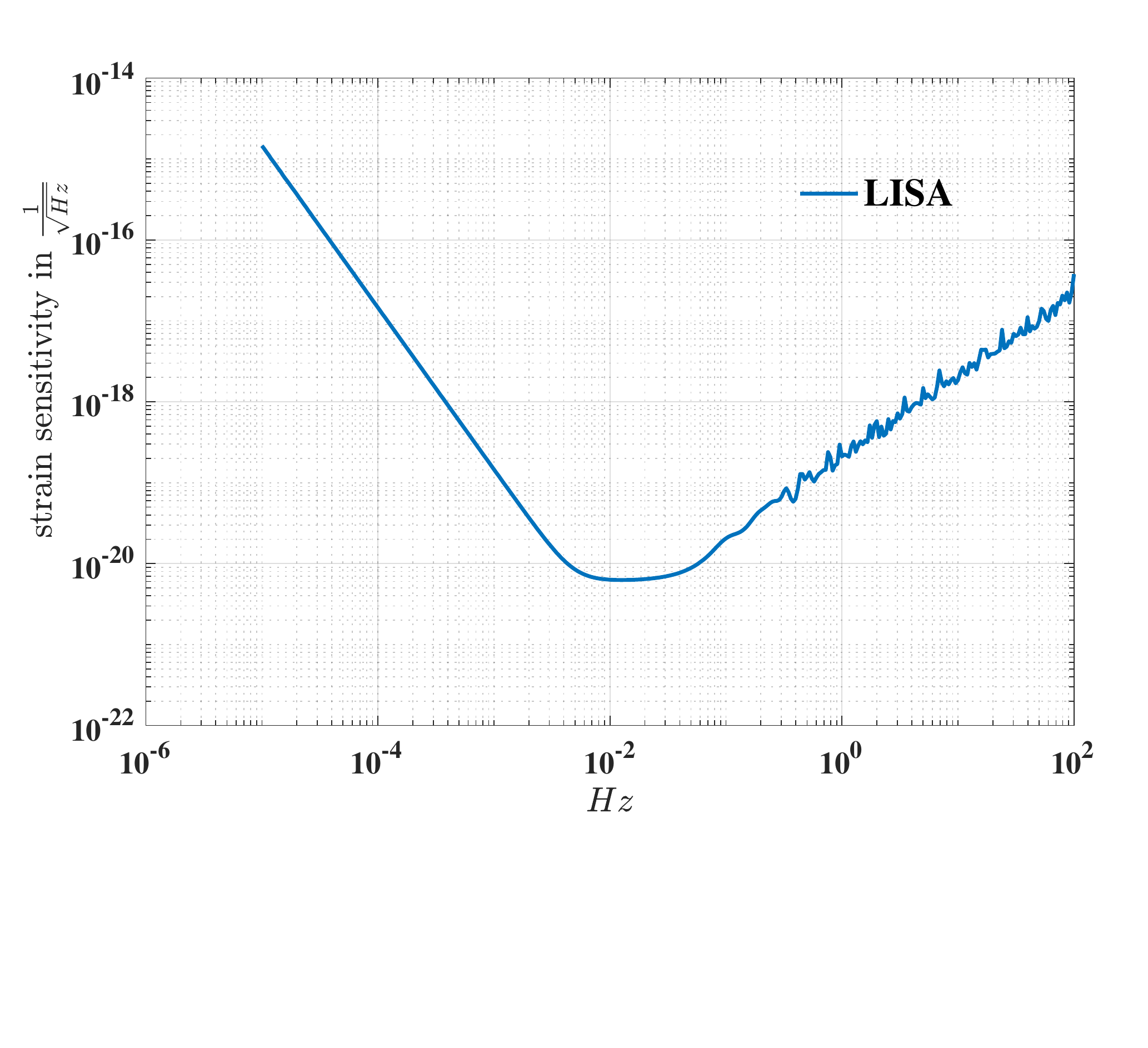}
\caption{The sensitivity curve of Taiji and LISA. The orange line
is the sensitivity curve of Taiji and the blue line is the
sensitivity curve of LISA.\label{fig:sensitivity}}
\end{figure*}

By assuming laser frequency noise is sufficiently suppressed in
TDI, the acceleration noise and optical path noise are considered
to be the dominant noises for GW observation. The budges of
acceleration noise for LISA and Taiji are considered as same
\cite{LISA:2017pwj,Yang:2022cgm},
\begin{equation}
 S_{\rm acc} = 9 \ \frac{\rm fm^2/s^4}{ \rm Hz } \left[ 1 + \left( \frac{0.4 \ {\rm mHz}}{f} \right)^2 \right]
 \left[ 1 + \left(\frac{f}{8 \ {\rm mHz}} \right)^4 \right] ,
\end{equation}
meanwhile, their optical path noise budgets are slightly different
as
\begin{align}
 S_{\rm op, LISA} & = 100 \ \frac{\rm pm^2}{\rm Hz} \left[ 1 + \left(\frac{2 \ {\rm mHz}}{f} \right)^4 \right],  \\
S_{\rm op, Taiji} & = 64 \ \frac{\rm pm^2}{\rm Hz} \left[ 1 + \left(\frac{2 \ {\rm mHz}}{f} \right)^4 \right].
 \end{align}
The power spectrum density (PSD) of noise which is presented in
Fig.~\ref{fig:sensitivity} is calculated by using the numerical
method, see Refs.~\cite{Wang:2020pkk,Wang:2020fwa} for detailed
algorithm.

\subsection{Data Curation}

\label{sec:data}

\begin{figure*}[ht]
\centering

\subfigure[$q=1$,
$m_1=2\times10^6M_{\odot}$,
$m_2=2\times10^6M_{\odot}$]{
\begin{minipage}[t]{0.5\linewidth}
\centering
\includegraphics[width=\columnwidth]{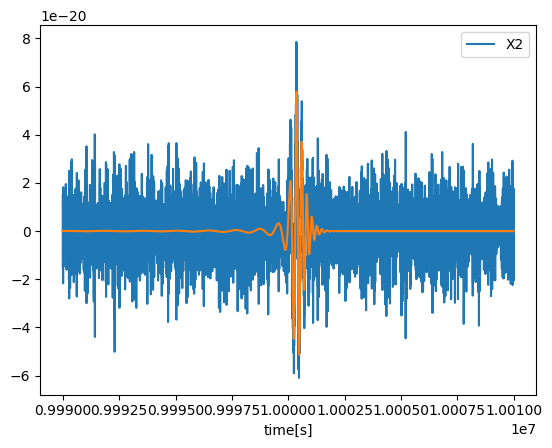}
\end{minipage}
}%
\subfigure[$q=1.5$,
$m_1=3\times10^6M_{\odot}$,
$m_2=2\times10^6M_{\odot}$]{
\begin{minipage}[t]{0.5\linewidth}
\centering
\includegraphics[width=\columnwidth]{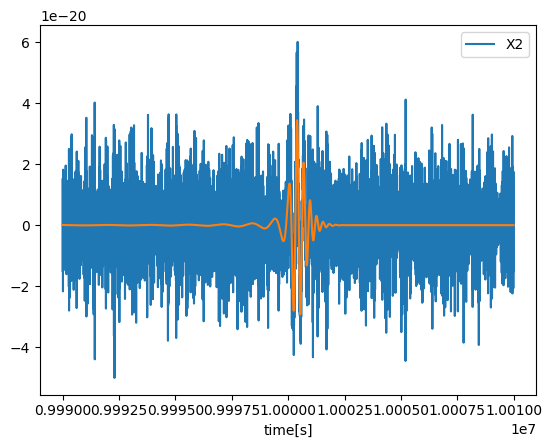}
\end{minipage}
}%

\subfigure[$q=2.5$,
$m_1=5\times10^6M_{\odot}$,
$m_2=2\times10^6M_{\odot}$]{
\begin{minipage}[t]{0.5\linewidth}
\centering
\includegraphics[width=\columnwidth]{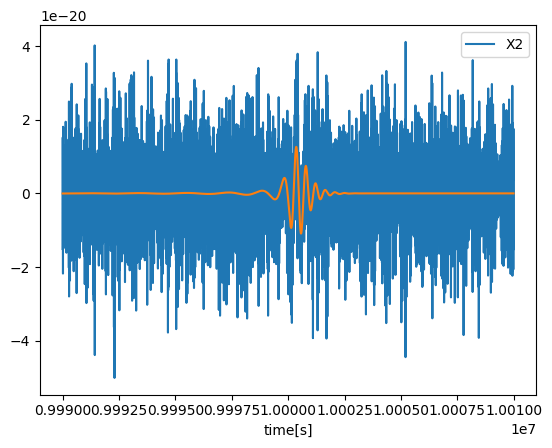}
\end{minipage}
}%
\subfigure[$q=5$,
$m_1=10^7M_{\odot}$,
$m_2=2\times10^6M_{\odot}$]{
\begin{minipage}[t]{0.5\linewidth}
\centering
\includegraphics[width=\columnwidth]{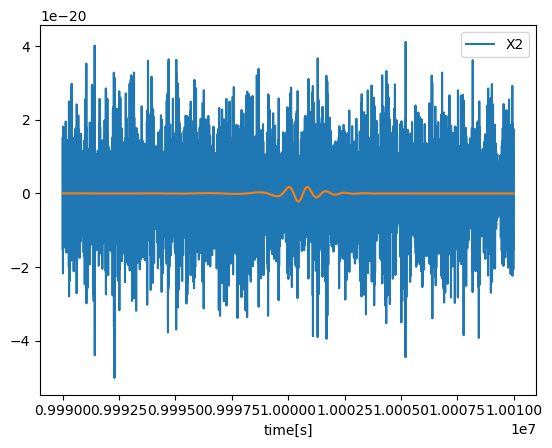}
\end{minipage}
}%

\centering \caption{Time domain representations of the TDI-X
channel with different mass ratio $\frac{m_1}{m_2}=q$. The data
from the LDC-2 dataset contain one MBHB signal presented by the
orange line, the noise+signal data is presented by the blue line.
\label{fig:LDC-X2}}
\end{figure*}

We used the signals extracted by the neural network to create new datasets for testing our model's ability to detect GW signals.
The test datasets consisted of 10,000 samples containing signals and 10,000 samples containing only noise. For the MBHB signals,
we set the SNR between 30 to 50, and the false alarm rate at  $1\%$.
For simulated MBHB signals, we used \texttt{SEOBNRv4\_opt}, which
is a version of the \texttt{SEOBNRv4} code \cite{Bohe:2016gbl}
with significant optimizations, which can bring the signals for a
high spin, high mass ratio MBHB system. We adopted the log-uniform
distribution for the parameter $M_{tot}$ from Ref.
\cite{Katz:2021uax}.To facilitate the visual understanding of the signal amplitude and SNR trends, we have presented a
time-domain illustration of the waveform in conjunction with the LDC dataset in Fig.~\ref{fig:LDC-X2} and the
detailed parameters range is shown in Table \ref{tab:mbhb_par}.

\begin{table}[htbp]
\begin{center}
\begin{minipage}{174pt}
\caption{Summary of parameter setups in MBHB signal simulation.}\label{tab:mbhb_par}%
\begin{tabular}{@{}ccc@{}}
\toprule
Parameter&Lower bound & Upper bound\\
\midrule
$M_{tot}$       & $10^6M_\odot$     & $10^8M_\odot$  \\
$q$             & $0.1$            & $10$    \\
$s_1^z$         & $-0.99$           & $0.99$ \\
$s_2^z$         & $-0.99$           & $0.99$ \\
$\phi_0$         & $0$           & $2\pi$ \\
$\iota$         & $0$           & $2\pi$ \\
$\delta$         & $0$           & $2\pi$ \\
\bottomrule
\end{tabular}
\end{minipage}
\end{center}
\end{table}

Hereafter, we project the signal to space-based detector,
and inject the projected signal to the noise with specific optimal
SNR as
\begin{align}
\label{eq:snr}
    {\rm SNR} = \left(s\mid s\right)^{-1/2},
\end{align}
Here, $s$ represents the signal, and $(h\mid s)$ is
\begin{equation}
(h \mid s) = 2\int_{f_{min}}^{f_{max}}
\frac{\tilde{h}^*(f)\tilde{s}(f)+\tilde{h}(f)\tilde{s}^*(f)}{S_n(f)}\,
df,
\end{equation}
where $f_{min}=3\times 10^{-5} {\rm Hz}$ and $f_{max}=0.05{\rm
Hz}$, and $S_n(f)$ is the noise PSD. Here, following the setting
of the LDC-2 dataset and Taiji configuration
\cite{Ruan:2018tsw,Ruan:2020smc}, we set the SNR to $20$\cite{Gair:2022knq}. Then the
data was whitened and normalized to $[-1,1]$. During the whitening
procedure, we applied the Tukey window with $\alpha=1/8$.

\section{Results}

Our GW twin design comprises two neural networks, Net A for LISA embedding and Net B for Taiji embedding. We concatenate the
embeddings and use them as input to the fully connected (FC) layer, where the weights of Net A and Net B are frozen. We only train
the FC layer using a subset of the training dataset. Our experiments show that training the FC layer for only 7-8 epochs in TDI-1.0
scenario is sufficient, and we found the same for TDI-1.5 and TDI-2.0 scenarios. Therefore, we choose to consider eight epochs for
all three generations.

We evaluate our approach based on five target metrics: AUC score, accuracy score, precision score, recall score, and F1 score. The
AUC score reflects the area under the receiver operating characteristic (ROC) curve, which plots the true positive rate (TPR) against
the false positive rate (FPR) for all classification thresholds.

\begin{equation}
TPR=\frac{TP}{TP+FN},
\end{equation}
where TP stands for``True Positive" and FN stands for``False
Negative",
\begin{equation}
FPR=\frac{FP}{FP+TN},
\end{equation}
where FP stands for``False Positive" and TN stands for``True
Negative".

\begin{figure}
\includegraphics[width=\columnwidth]{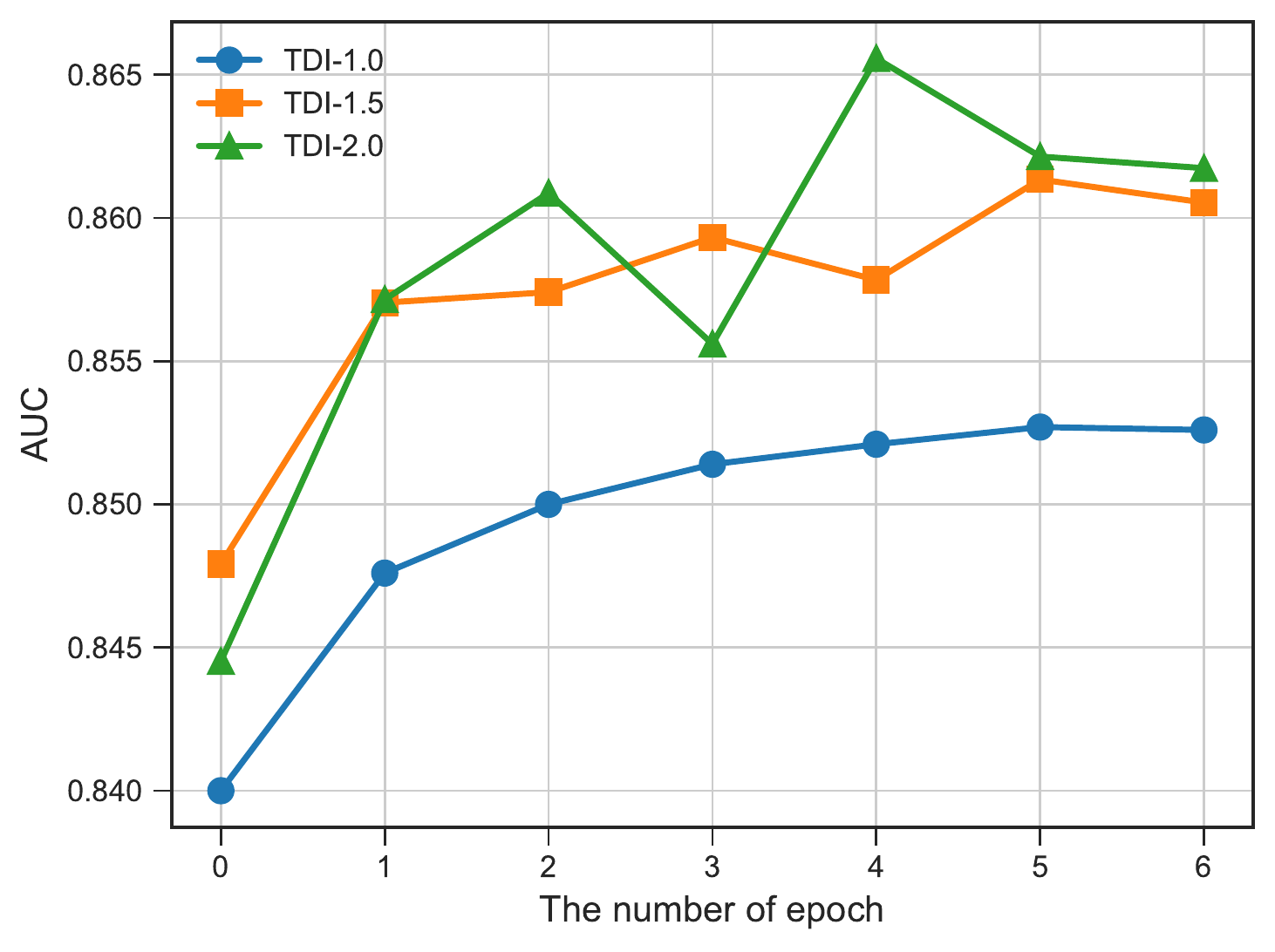}
\caption{AUC with respect to epoch. The vertical axis represent the value of AUC for SSL model aim at detecting MBHB and the horizontal axis
represent the sequence label of epoch.\label{fig:AUC}}
\end{figure}

\begin{figure}
\includegraphics[width=0.774\columnwidth]{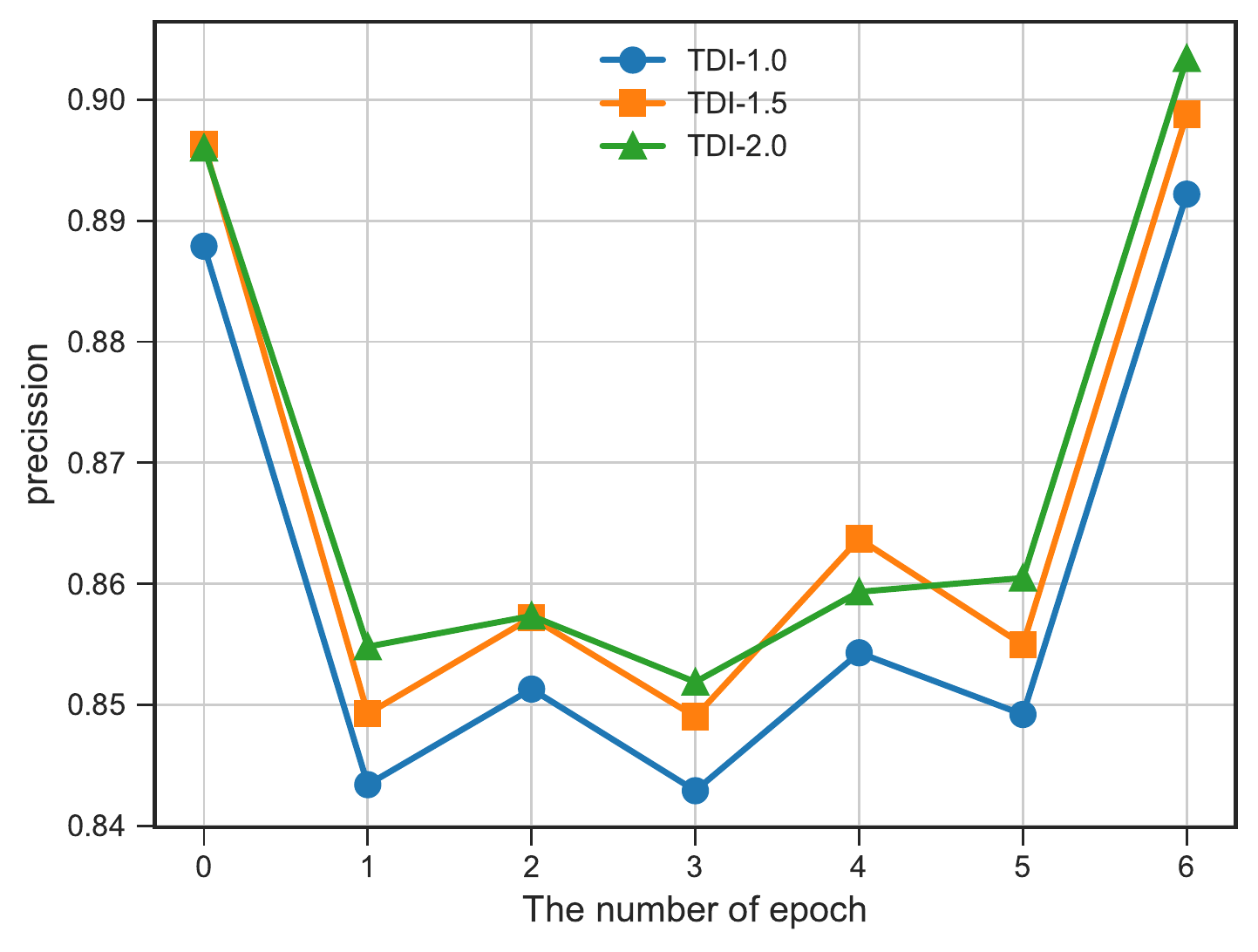}
\includegraphics[width=0.774\columnwidth]{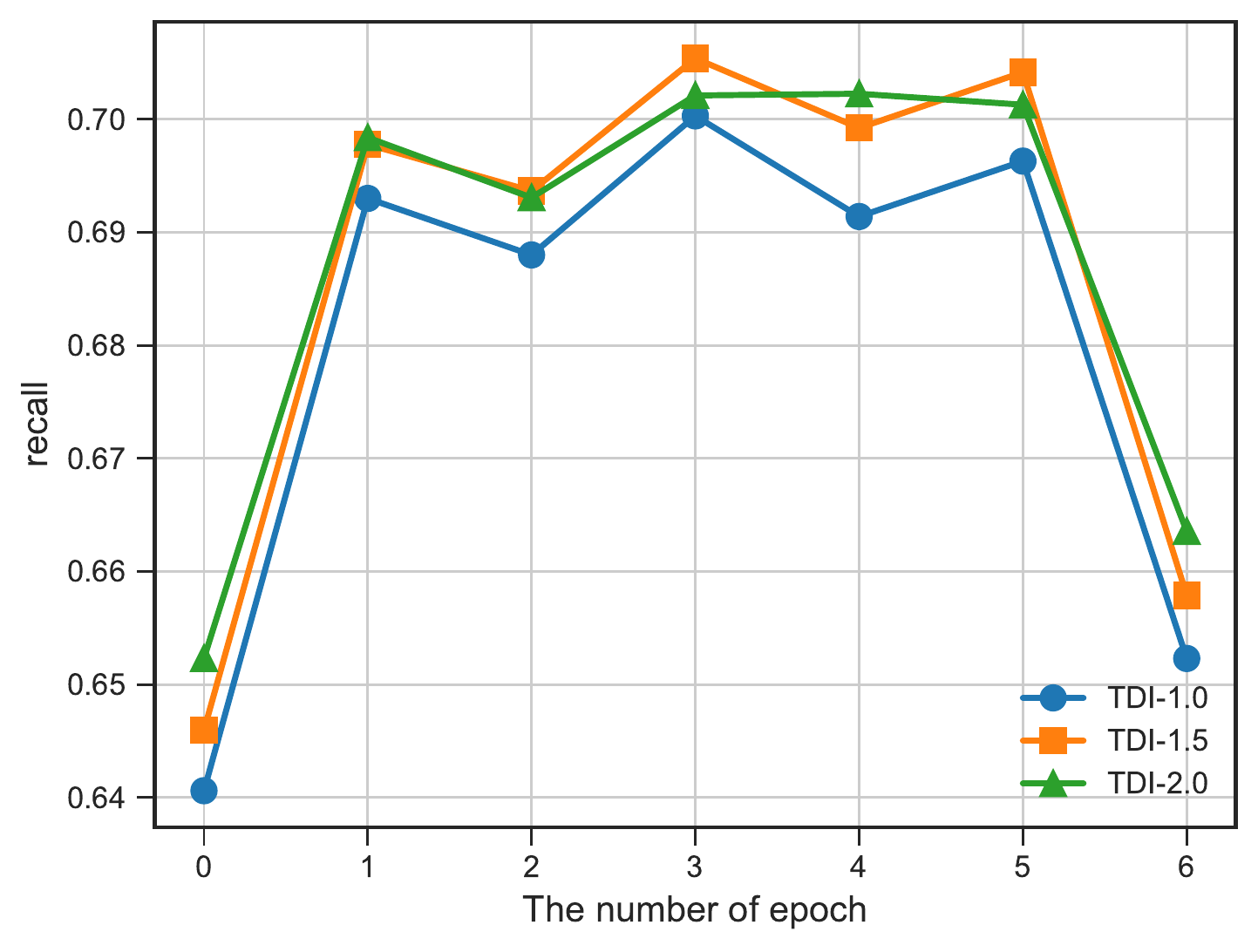}
\caption{Precision and Recall with respect to
epoch under the TDI-1.0(blue), TDI-1.5(orange) and TDI-2.0(green). The vertical axis of the both subfigure
 represent the value of precision score and the value of recall,  the horizontal axis of the both subfigure represent the sequence label of epoch. \label{fig:PR}}
\end{figure}

\begin{figure}
\includegraphics[width=0.774\columnwidth]{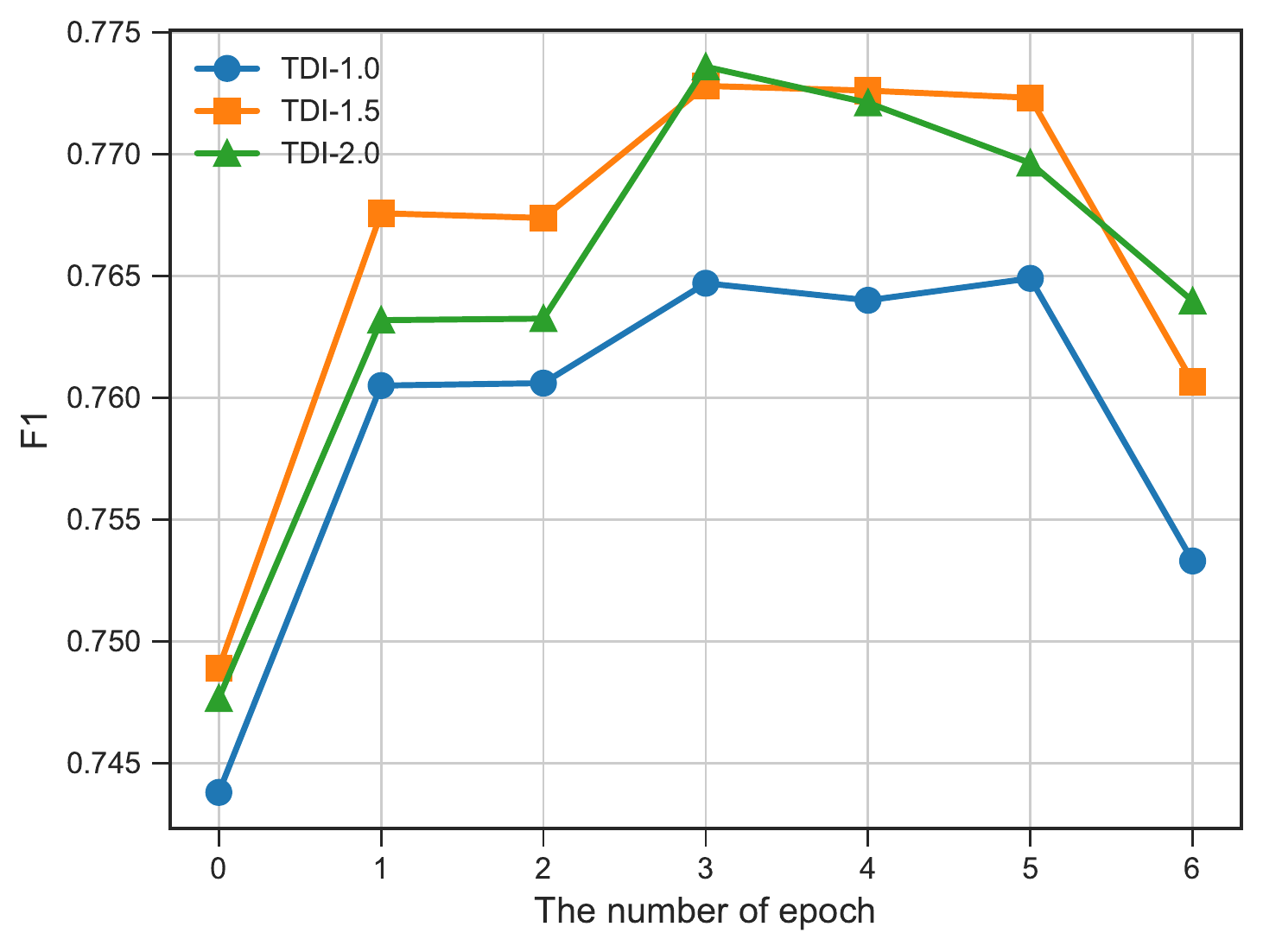}
\includegraphics[width=0.774\columnwidth]{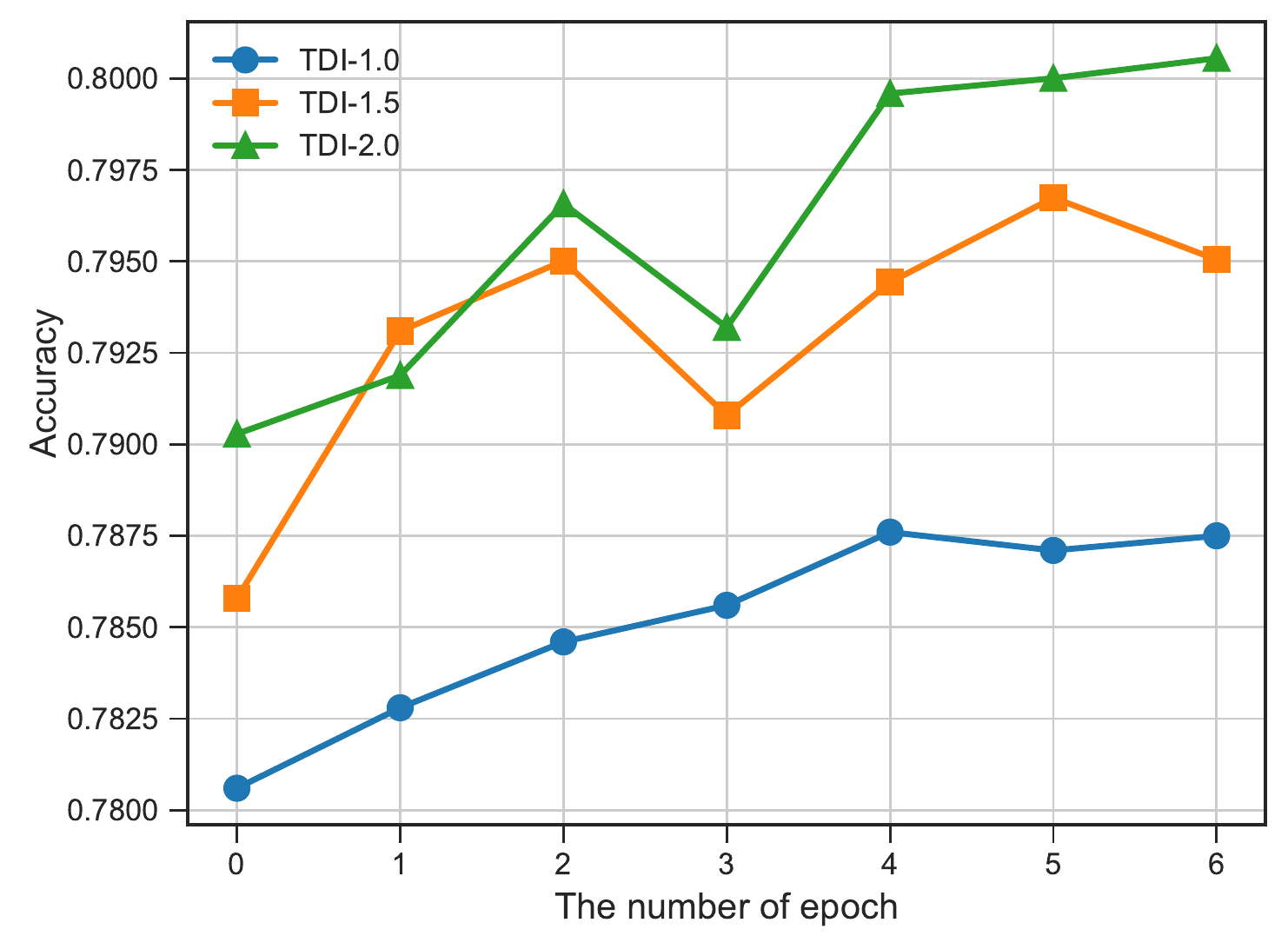}
\caption{F1 score and accuracy with respect to epoch under the TDI-1.0(blue), TDI-1.5(orange) and TDI-2.0(green). The vertical axis of the both subplotting
 represent the value of F1 score and the value of accuracy,  the horizontal axis of the both subplotting represent the sequence label of epoch. \label{fig:FA}}
\end{figure}

Fig.\ref{fig:AUC} shows the evolution of AUC values with increasing epoch numbers across TDI-1.0, TDI-1.5, and TDI-2.0 generations. The maximum AUC value ($0.85271$) occurs in the 7th epoch of the TDI-1.0 generation. Interestingly, the AUC value for the 8th epoch is slightly smaller than the 7th epoch, which may be due to overfitting caused by the $1\times1$
 convolution layer between the two epochs. Our results suggest that our self-supervised learning model's ability to extract MBHB GW signals is sufficient after only 7 epochs of training. In TDI-1.5 and TDI-2.0, the AUC value increases monotonically with an increasing number of training epochs. Notably, the AUC value for TDI-2.0 is greater than TDI-1.5, indicating that our SSL model is more accurate under more realistic TDI assumptions.
\begin{figure*}[htbp]
\centering
\subfigure[JS divergence for $M_1$ and $M_2$ marginalization posterior between different TDI generation assumptions with LISA MDC data.]
{
    \begin{minipage}[b]{.45\linewidth}
        \centering
        \includegraphics[scale=0.1]{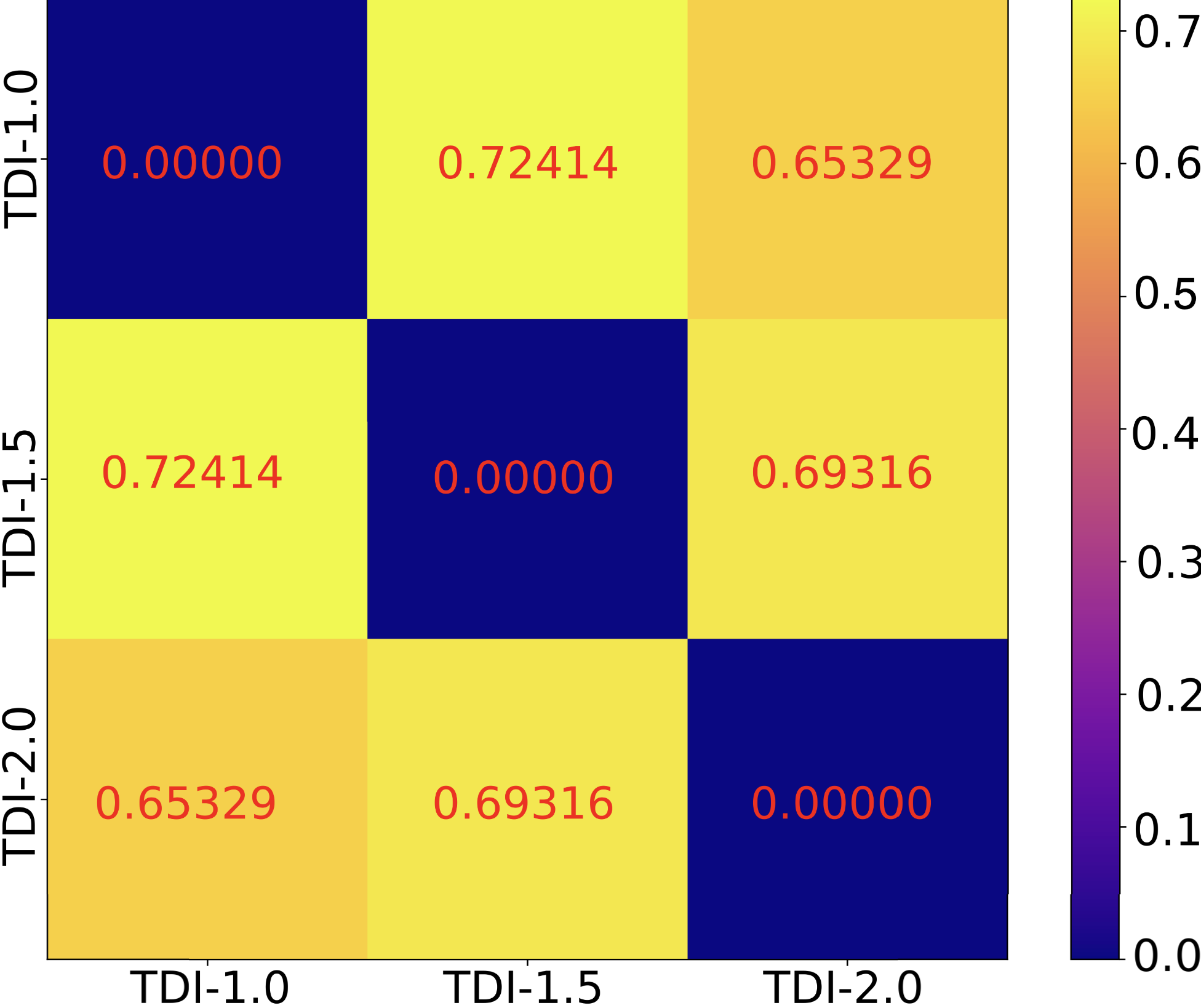}
    \end{minipage}
    \begin{minipage}[b]{.45\linewidth}
        \centering
        \includegraphics[scale=0.1]{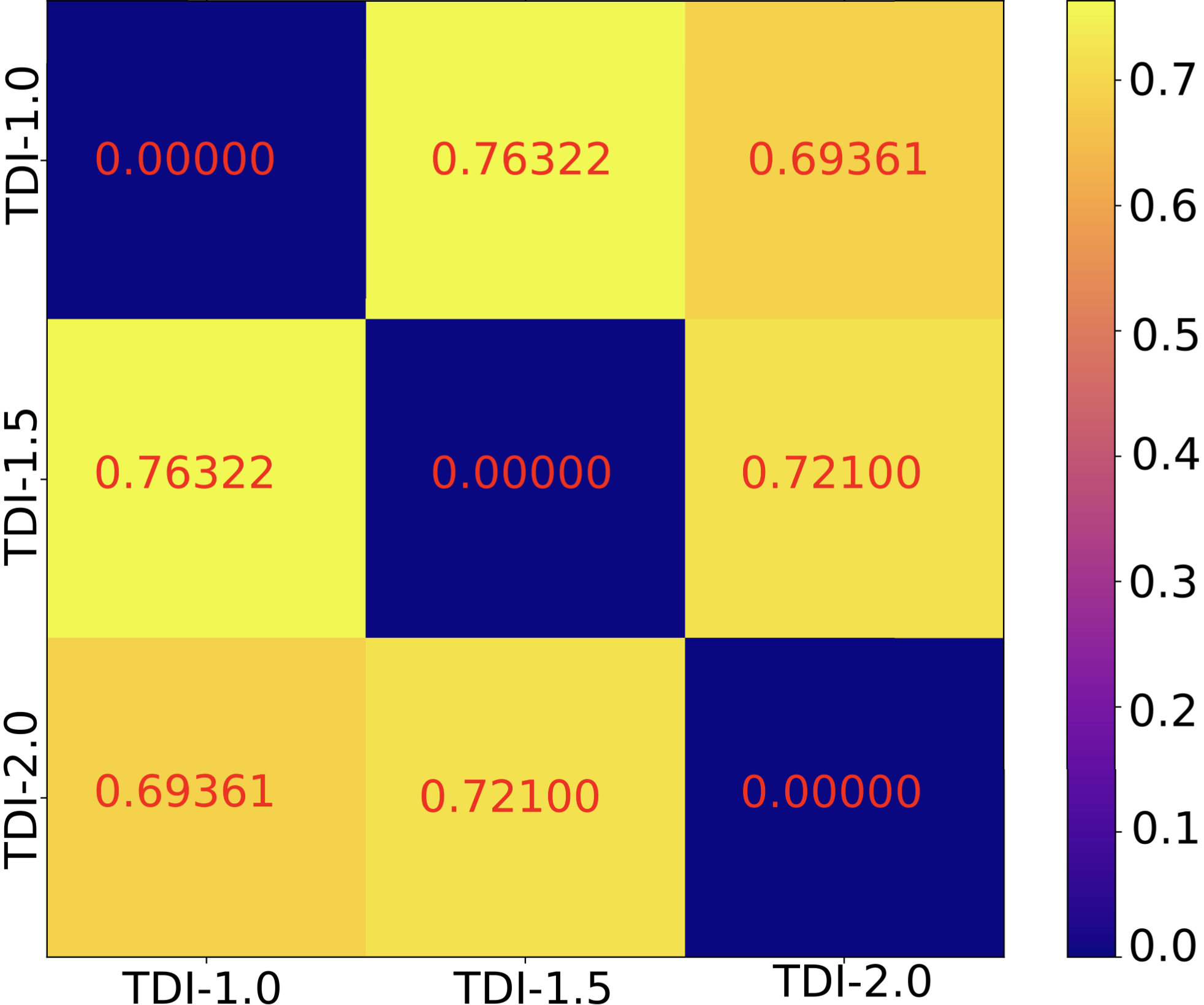}
    \end{minipage}
}
\subfigure[JS divergence for $M_1$ and $M_2$ marginalization posterior between different TDI generation assumptions with Taiji MDC data.]
{
    \begin{minipage}[b]{.45\linewidth}
        \centering
        \includegraphics[scale=0.1]{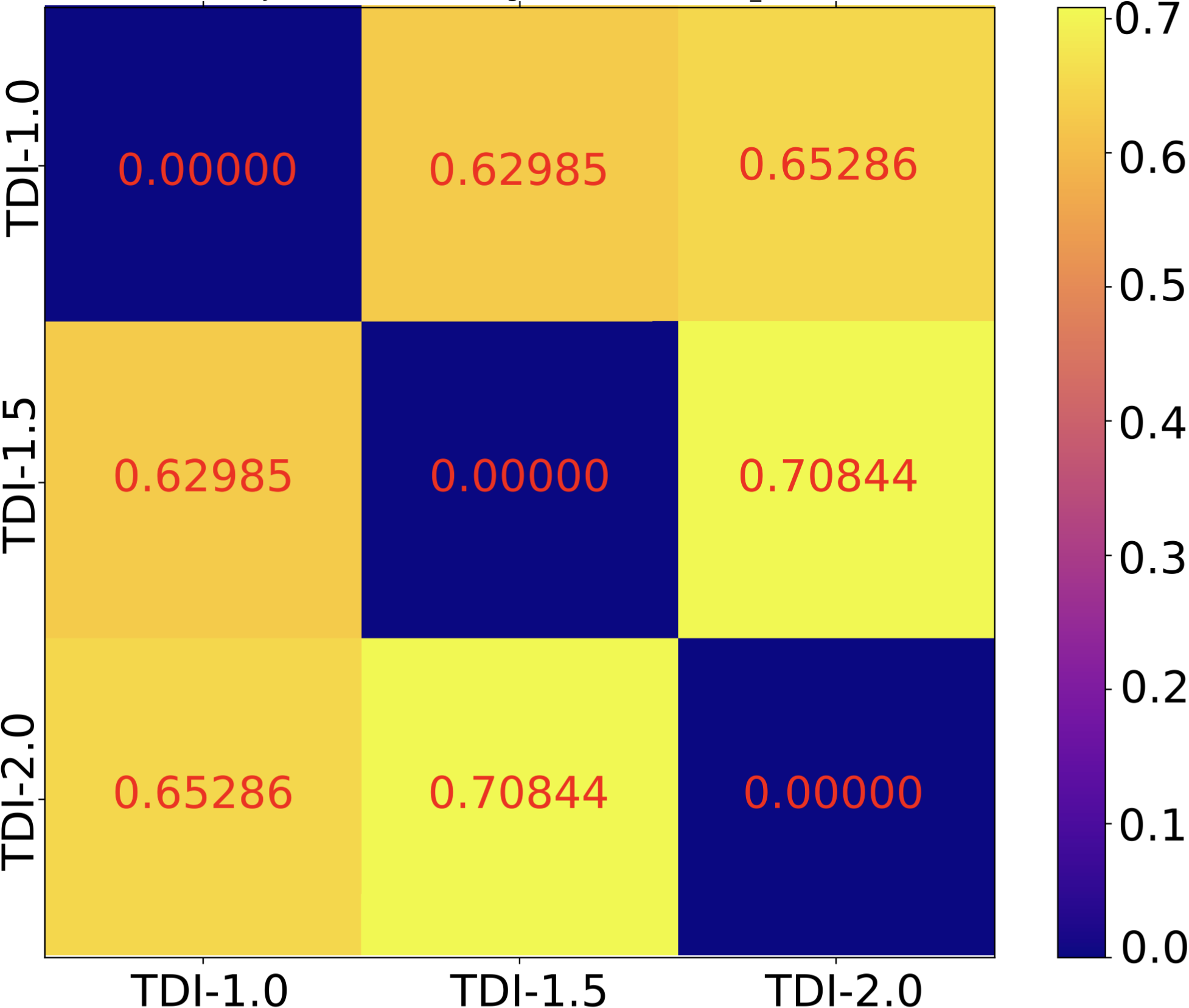}
    \end{minipage}
    \begin{minipage}[b]{.45\linewidth}
        \centering
        \includegraphics[scale=0.1]{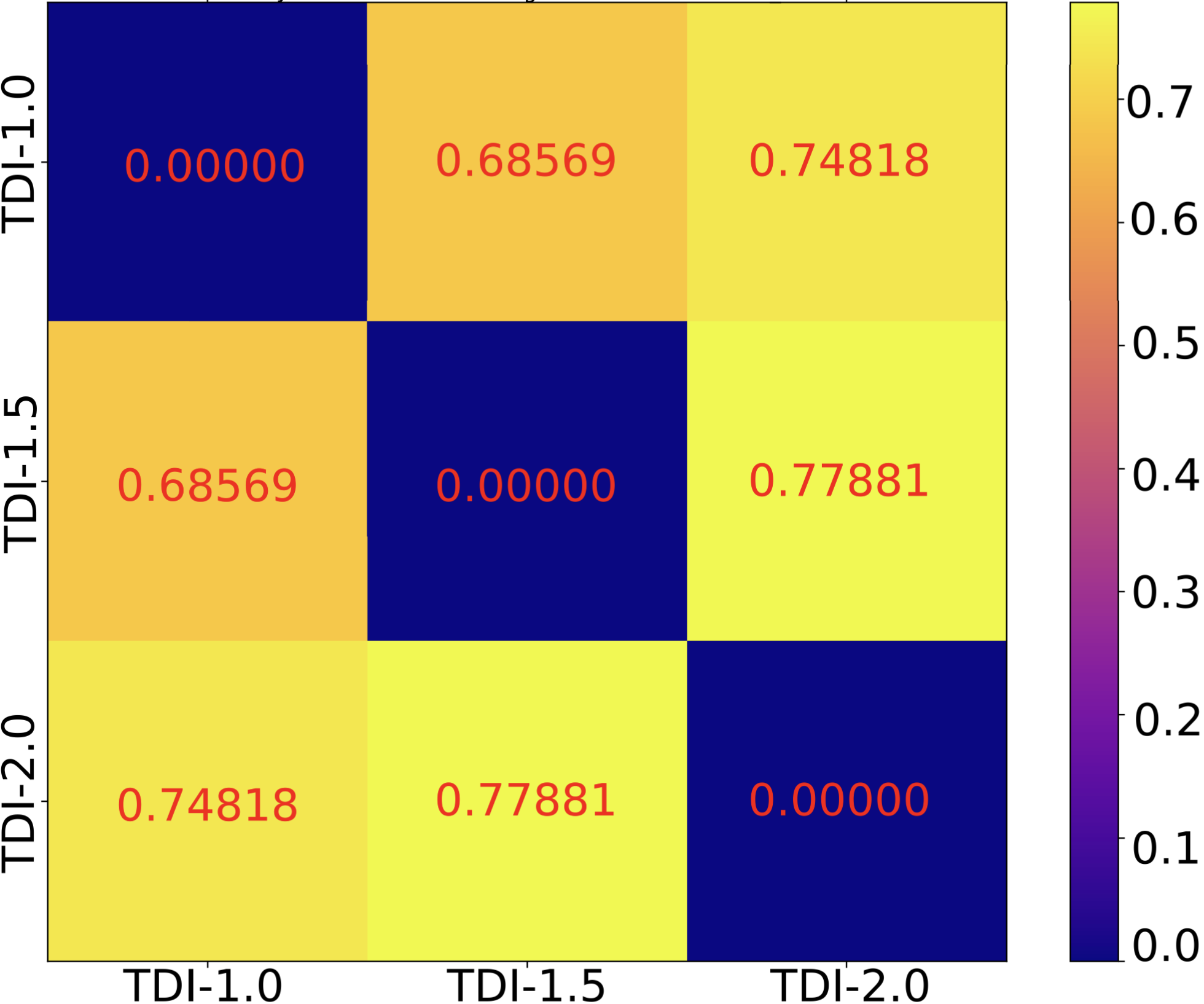}
    \end{minipage}
}
\caption{JSD for primary mass $M_1$ and secondary mass $M_2$ marginalization posterior between different TDI generation assumption with LISA and Taiji MDC data.\label{fig:mass}}
\end{figure*}

\begin{figure*}[htbp]
\centering
\subfigure[JS divergence for $\phi$, $\iota$ and $\delta$ marginalization posterior between different TDI generation assumptions with LISA MDC data.]
{
    \begin{minipage}[b]{.3\linewidth}
        \centering
        \includegraphics[scale=0.068]{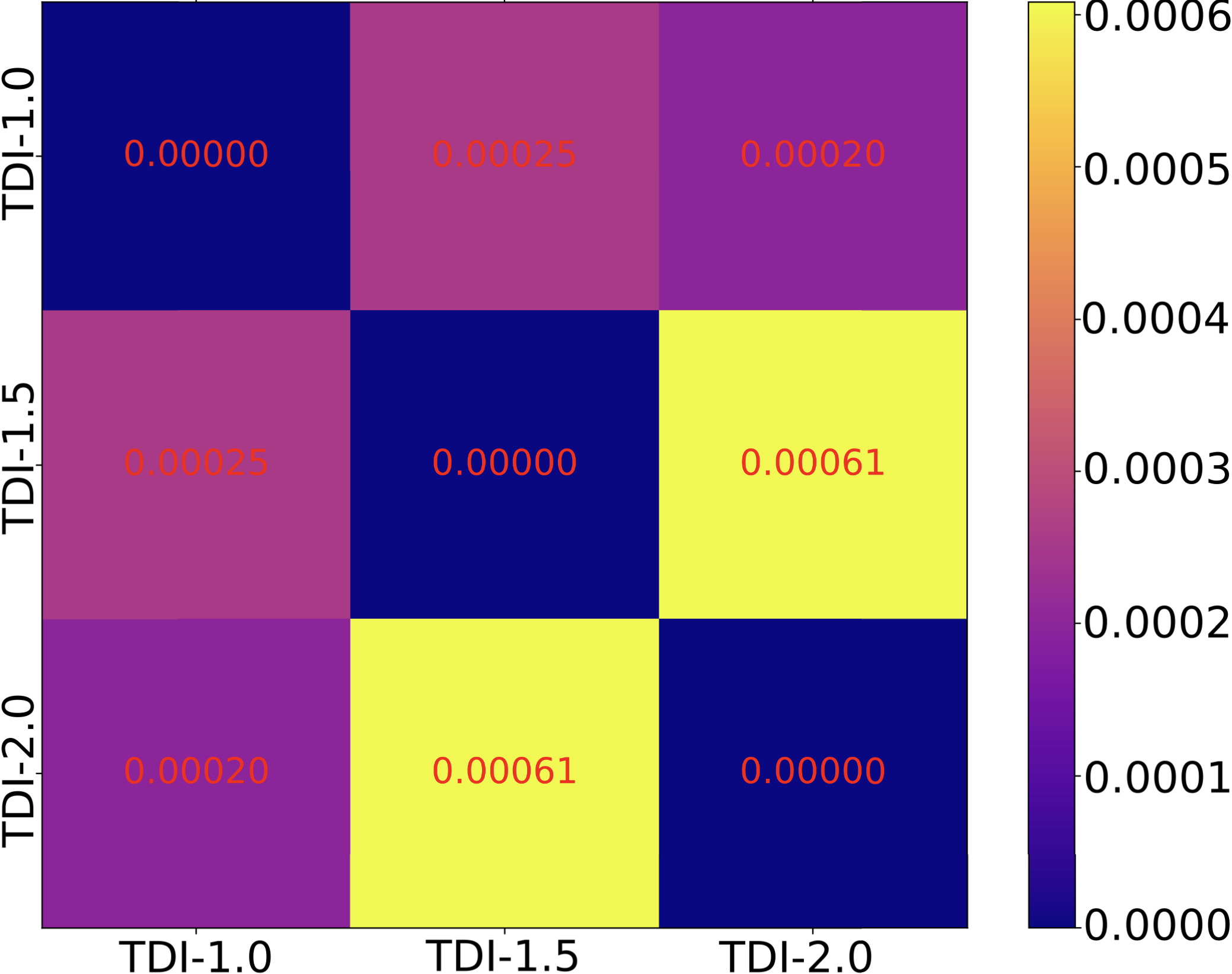}
    \end{minipage}
    \begin{minipage}[b]{.3\linewidth}
        \centering
        \includegraphics[scale=0.068]{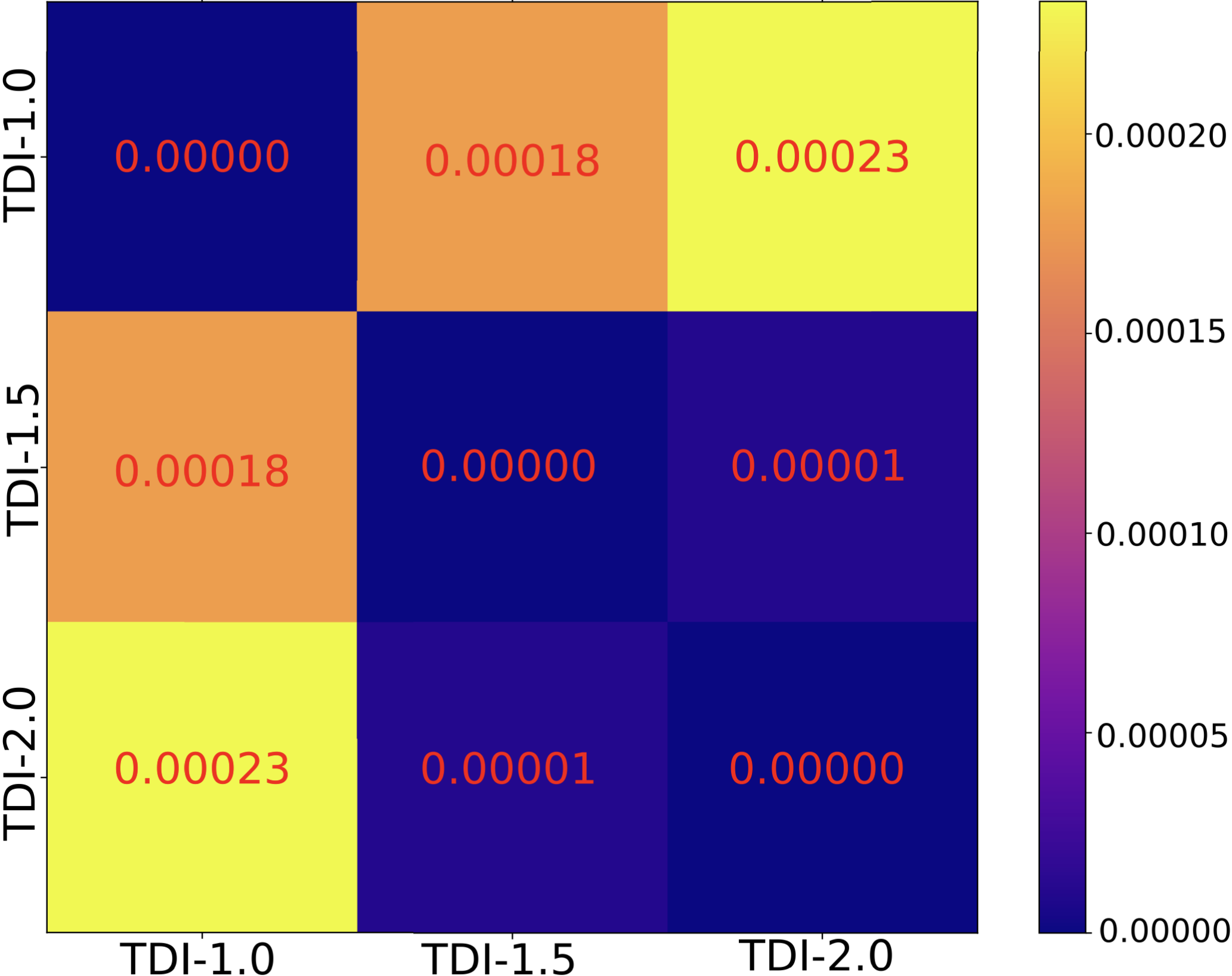}
    \end{minipage}
    \begin{minipage}[b]{.3\linewidth}
        \centering
        \includegraphics[scale=0.068]{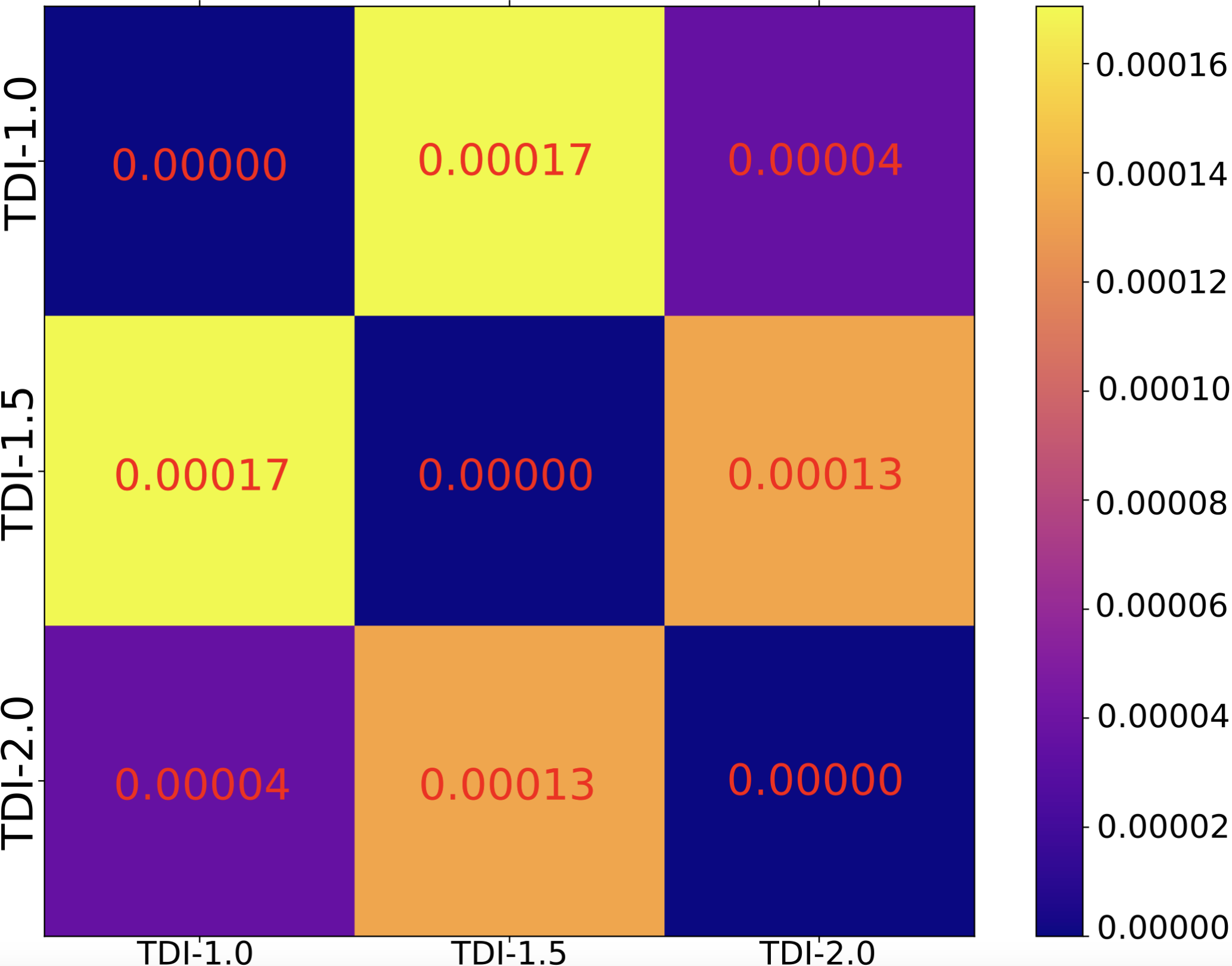}
    \end{minipage}
}
\subfigure[JS divergence for $\phi$, $\iota$ and $\delta$ marginalization posterior between different TDI generation assumptions with Taiji MDC data.]
{
    \begin{minipage}[b]{.3\linewidth}
        \centering
        \includegraphics[scale=0.068]{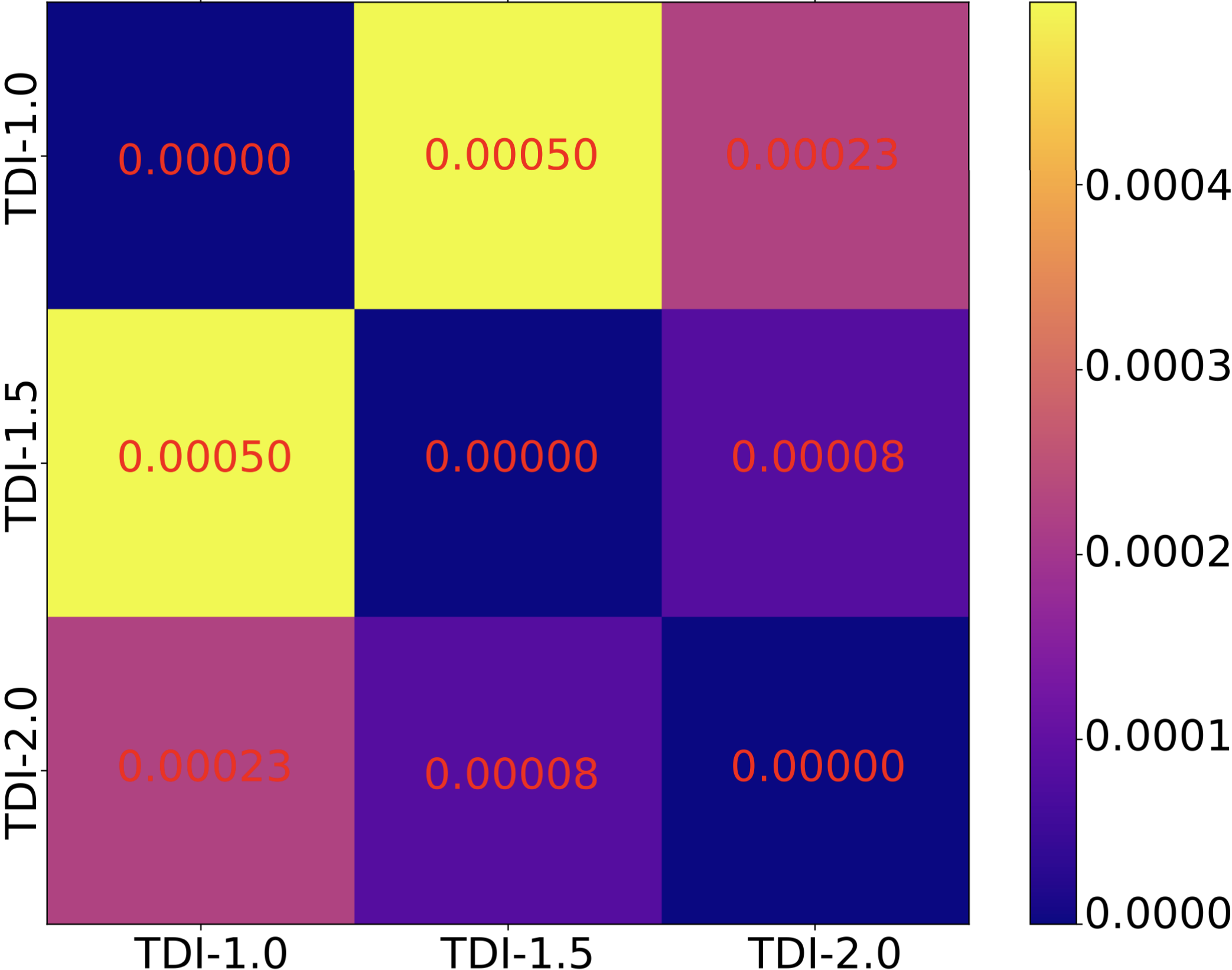}
    \end{minipage}
    \begin{minipage}[b]{.3\linewidth}
        \centering
        \includegraphics[scale=0.068]{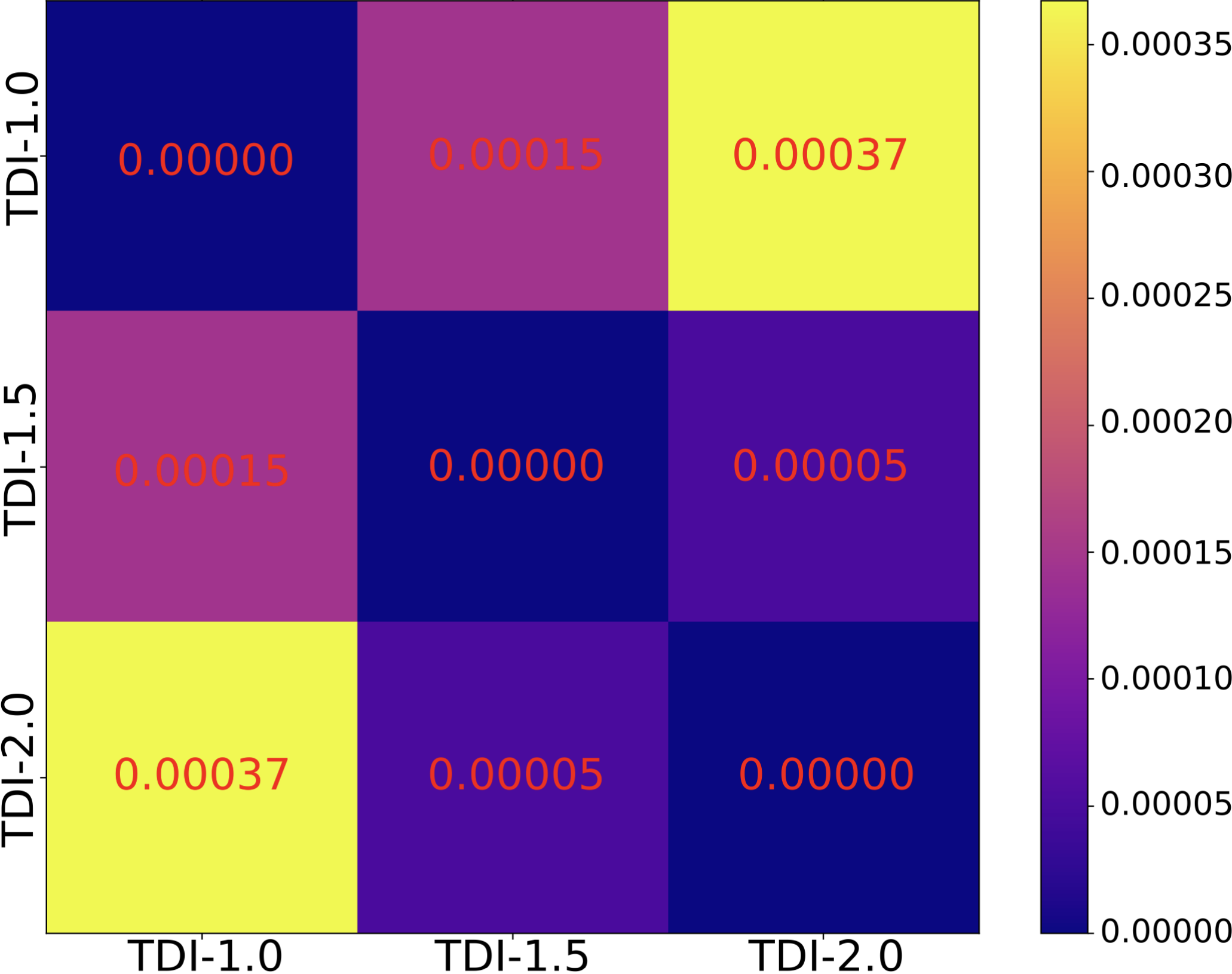}
    \end{minipage}
    \begin{minipage}[b]{.3\linewidth}
        \centering
        \includegraphics[scale=0.068]{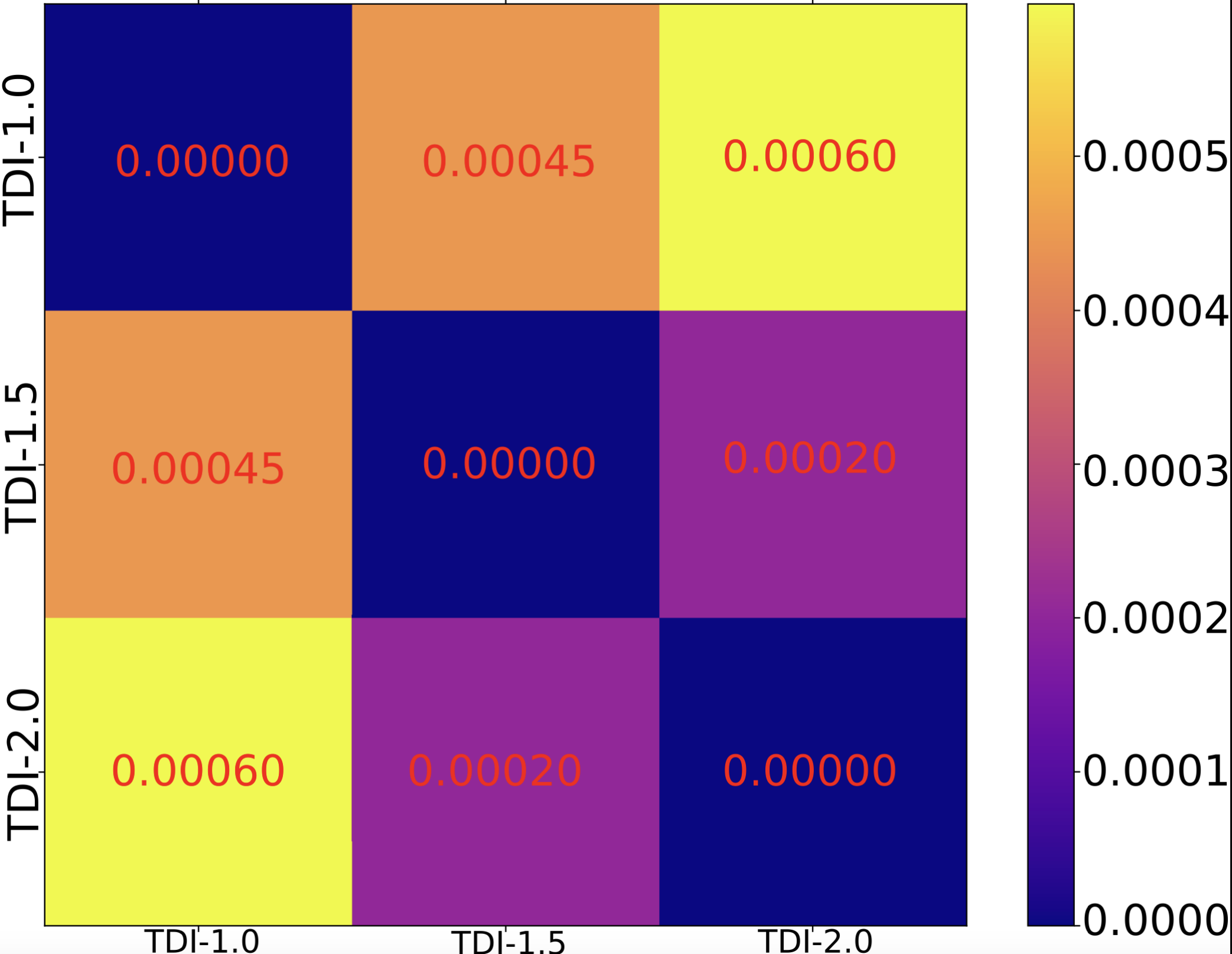}
    \end{minipage}
}
\caption{JSD for phase $\phi$, inclination angle $\iota$ and polarization angle $\delta$ marginalization posterior between different TDI generation assumption with LISA and Taiji MDC data.\label{fig:skymap_PE}}
\end{figure*}

We also show the precision and recall. The Precision attempts to
describe the proportion of actually correct positive
identifications,
\begin{equation}
Precision=\frac{TP}{TP+FP}.
\end{equation}
The Recall describes the proportion of identified correctly actual
positives,
\begin{equation}
Recall=\frac{TP}{TP+FN},
\end{equation}

Our model achieves high Precision scores in the 7th epoch, suggesting correct detection of GW signals by our self-supervised learning model. The relatively high Recall rate indicates a relatively low rate of false negatives; however, precision and recall are often in tension such that improving one typically reduces the other. To evaluate the effectiveness of our model, we analyze both precision and recall, and utilize the F1 score to provide a holistic assessment in signal retrieval. The F1 score is a frequently-used metric in machine learning that evaluates the performance of binary classification models by balancing precision and recall.
We define the F1 score using the equation:
\begin{equation}
F1=\frac{2TP}{2TP+FP+FN},
\end{equation}

 In addition to the F1 score, we also calculate Accuracy, a widely used evaluation metric for classification models. Informally, accuracy measures the fraction of correct predictions made by our model and is defined as:
 \begin{equation}
Accuracy=\frac{TP+TN}{TP+TN+FP+FN}.
\end{equation}

In Figure \ref{fig:FA}, we show the changes in F1 score and accuracy with respect to epoch number. Both metrics steadily increase, indicating that our self-supervised learning (SSL) approach supports good generalization for Massive Binary Black Hole (MBHB) signal searching under any Time Delay Interferometry (TDI) generation assumption, from a holistic perspective. We also observe that the F1 score and accuracy levels are similar, suggesting that our SSL model is capable of handling imbalanced data, which is a significant advantage of SSL.

We investigate the effect of different TDI generations on the SSL method's performance in estimating five key parameters (
$M_1$, $M_2$, $\phi$, $\iota$, $\delta$), using a modified DINGO-GW code package\cite{GW:DINGO} based on the neural parameter estimation (NPE) method. Our analysis uses data that we trained with our SSL method, and we evaluate the influence of different TDI generations using the Jenson-Shannon (JS) divergence metric. Smaller JS divergence values indicate that parameter estimations for different TDI generations are comparable, supporting our inference that the SSL method is minimally influenced by the TDI generation assumptions. As seen in Figure.\ref{fig:mass}, the influence of different TDI generation assumptions on the primary and secondary mass is significant in both the LISA and Taiji MDC datasets. However, as Figure.\ref{fig:skymap_PE} shows, the effect on $\phi$, $\iota$ and $\delta$
 in both datasets is negligible. This is because the SNR of the MBHB signal is most strongly influenced by changes in primary and secondary mass, while the contribution from phase and skymap localization angle is minor. Therefore, despite the influence of TDI generation assumptions on mass, the SSL method's performance in estimating $\phi$, $\iota$ and $\delta$ remains consistent across different TDI generations.

\begin{figure*}
\includegraphics[width=13.7cm]{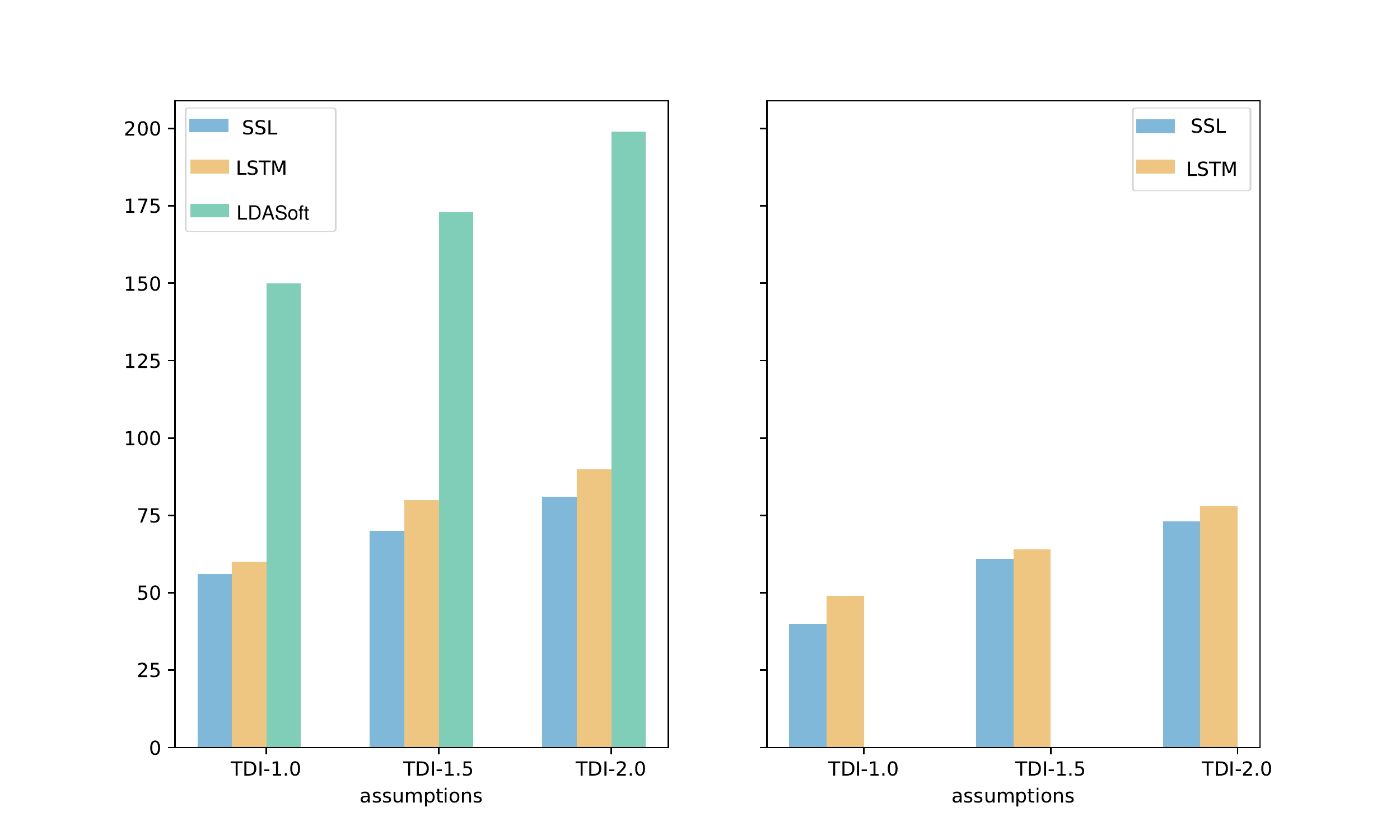}
\caption{We compare the time cost for three methods - SSL (blue histogram), LSTM (yellow histogram), and LDASoft (green histogram) - for different TDI generations. The left plot displays the results for LISA MDC data, where we compare the time cost of the three main methods. The right plot shows the results for TAIJI MDC data, where we compare the time cost of both machine learning methods.\label{fig:time_cost}}
\end{figure*}

SSL also offers significant computational efficiency advantages over other methods. To demonstrate this, we compare SSL to the Long Short-Term Memory network (LSTM)\cite{Hochreiter:1997yld} and LDASoft\cite{LDA:LDASoft} methods on the LISA and Taiji MDC datasets. Using all three methods, we search for specific injection signals and evaluate the time costs of each. Figure \ref{fig:time_cost} shows that LDASoft incurs the highest time cost among the three methods in the LISA MDC data environment, steadily increasing as TDI configuration approaches the true value. Conversely, SSL and LSTM significantly reduce time costs by over 70\% as compared to LDASoft, with SSL being the most efficient. The annotation costs and overall computational costs are also lower with SSL. We believe that further methodological enhancements and the availability of more robust training organizations will make SSL even more promising for controlling computational costs.

\section{Conclusion}

In this paper, we present a self-supervised learning model for GW detection. Our model is based on simulated massive black hole binary signals embedded in synthetic Gaussian noise that matches space-based GW detectors' Taiji and LISA sensitivity. Our results demonstrate that the self-supervised learning model we propose is capable of extracting signals and that it is a highly efficient method for GW signal identification.

To accomplish this, we utilize a corresponding contrastive learning method in which we propose the GW twins design operating on a joint embedding of augmentation vectors (where Net A and Net B represent the LISA and Taiji embeddings, respectively). This design is conceptually simple, easy to implement, and benefits from high-dimensional embeddings that do not require large batches, asymmetric mechanisms such as momentum encoders or non-differentiable operators, or stop-gradients.

We apply our SSL method to a simple signal search scenario in which we search for MBHB signals in detecting data. Our results show that the SSL method does well in recovering the injected MBHB signals from the Gaussian noise with both the LISA MDC and Taiji data settings. From various metric results, we conclude that SSL is more effective with a more realistic TDI generation assumption from an entire angle, and the different TDI generations can influence the primary mass and secondary mass inference results obviously and slightly for $\phi$, $\iota$ and $\delta$ when using the SSL method. The time cost of SSL is advantageous to LDASoft based on globalfit, and it also uses less computation cost than the LSTM method. It will reduce the computation cost and time with future optimizations.

Although we applied our method to the MBHB GW events, it could still be applied to other cases, such as binary white dwarf and extreme mass ratio inspirals. It might be helpful in searching for the GW echo signals, particularly the unequal interval echoes \cite{Wang:2018mlp,Wang:2018cum,Li:2019kwa}. The search for signals in GW data is mainly affected by non-Gaussian noise. Actually, our method can be applied in realistic non-Gaussian data. Note that the claim of matched-filtering optimality only applies in the Gaussian noise case. We will study these interesting issues in later works.

It is worth mentioning that we take Taiji and LISA as GW twins to present our method; however, self-supervised learning can also be applied in land-based GW detectors with high-quality data\cite{LIGO:2021ppb,Virgo:2022fxr,Virgo:2022ysc}.

\section{Acknowledgments}

WYT is supported by National Key Research and Development Program of China Grant No. 2021YFC2203004. We acknowledge the Tianhe-2 supercomputer for providing computing resources.

\end{document}